\documentclass[twocolumn,showpacs,aps,prl]{revtex4}
\usepackage{graphicx}
\usepackage{epsfig}
 \begin{document}
\setlength{\unitlength}{1mm}
\bibliographystyle{unsrt} 
\title{A linear Stark shift in dressed atoms as a signal to measure a nuclear anapole moment with a
cold atom fountain or interferometer}
\author{Marie-Anne Bouchiat  }
\affiliation{ Laboratoire Kastler Brossel, D\'epartement de Physique de l'Ecole
Normale Sup\'erieure,\\ 24 Rue Lhomond, F-75231  Paris  Cedex 05, France }
\date {November 3, 2006 }
 \newcommand \be {\begin{equation}}
\newcommand \ee {\end{equation}}
 \newcommand \bea {\begin{eqnarray}}
\newcommand \eea {\end{eqnarray}}
\newcommand \nn \nonumber
\def \(({\left(}
\def \)){\right)}
 \def \vs{{\mathbf{s}}}    
 \def \vI{{\mathbf{I}}}
 \def \vr{{\mathbf{r}}}
 \def \vF{{\mathbf{F}}}
 \def \vr{{\mathbf{r}}}
\def \vE{{\mathbf{E}}}
\def \vB{{\mathbf{B}}}
\def \ve{{\mathbf{e}}}
\def \vk{{\mathbf{k}}}
\def \vD{{\mathbf{D}}}
\def\bra{\langle}
\def\ket{\rangle}

\begin{abstract}
We demonstrate theoretically the existence of a linear dc Stark shift of the individual substates of an alkali atom
in its ground state, dressed by a  circularly polarized laser field. It arises from the electroweak nuclear
anapole moment violating P but not T. It is characterized by the pseudoscalar $\xi \vk
\wedge \vE \cdot \vB$ involving the photon angular momentum and static electric and magnetic
fields. We derive the relevant left-right asymmetry with its complete signature  
%its $F,m_F$-dependence, its magnitude and complete signature 
in a field configuration selected for  
a precision measurement with cold atom beams. The $3, 3 \rightarrow 4, 3$ Cs transition
frequency shift amounts to 7 $\mu$Hz for a laser power of  $\approx$ 1 kW at 877 nm, E=
100 kV/cm and B$\gtrsim$ 0.5 G. % We discuss the  feasibility of such an experiment. 
 \end{abstract}
\maketitle
 Atomic parity violation has built a bridge between atomic physics and both particle and nuclear physics
\cite{bou74}. The concept of a nuclear anapole moment
\cite{zel59} may be ascribed to toroidal currents \cite{fla04}. Alternatively, it describes a chiral distribution of
the atomic nucleus magnetization, {\it i.e.} a nuclear  helimagnetism,  resulting from the parity violating (PV)
nuclear forces exerted inside a stable nucleus \cite{CB91}. In a stationary atomic state, as a result of
T-reversal invariance, only the nuclear spin dependent PV interaction discussed here 
contributes to observable PV effects, 
 while in forbidden optical transitions the dominating
nuclear spin-independent PV interaction, associated with the weak nuclear charge $Q_W$, makes the anapole
effect difficult to extract.  

Today, there exists only one empirical manifestation of this nuclear anapole moment, though an
important nuclear property.  It arises from PV 
measurements performed on the highly forbidden 6S-7S Cs transition,  yielding electric dipole transition
amplitudes,
$E_1^{pv} \simeq 0.8\times 10^{-11} ea_0$, of a slightly different magnitude ($\pm 2\%$) for the two
hyperfine components
$\Delta F= \pm 1$ \cite{woo97}. The reported result presents significant discrepancies
with different theoretical predictions and with other manifestations of the PV nuclear forces \cite{fla04,CB91}. 
% in particular the circular polarizatin of photons in the decay of $^{18}$F  \cite{fla04,CB91} 
On the experimental side, it is notoriously difficult to extract such a small
fractional difference from  measurements where the control of systematic effects represents a major task.  In
short, there is a clear need for new independent results both in atomic and nuclear physics. In this latter field,
new high precision experiments are already underway, namely the measurement of the directional gamma-ray
asymmetry in the n+p $\rightarrow$ d+$\gamma$ experiment and of the neutron spin rotation in helium
\cite{snow}. We suggest here a new atomic physics approach using a frequency shift measurement in an
atomic fountain or a matter-wave interferometer.    

 Like previous proposals \cite{lov,ezh,deM,bou01}, this project is well
suited to an atom in its ground state, where the effect of the dominant PV potential due to the
weak charge cancels,  
but it presents important additional advantages. 
First, instead of relying on the left-right asymmetry in a transition rate, it is based, as in \cite{bou01}, on an 
energy shift of  the atomic states, and its measurement takes advantage
of the methods developed for the primary $^{133}$Cs standard. This experimental domain and that of
atomic interferometry \cite{bor,Chu} have now reached an impressive, continuously improving, level of
accuracy. Nowadays, beyond their  metrological interest, cold atom fountains and interferometers are becoming 
undisputable tools, leading to a wide variety of high precision measurements of fundamental interest, such as a
lower limit on the  time variation of the fine structure constant
\cite{Syr02}, new tests of Lorentz invariance (LI) \cite{cla}, possibly tighter constraints on the electron EDM
\cite{gou06,hin} and new tests of the Einstein Equivalence Principle
\cite{Syr02}. Second, this {\it anapole shift} is both a light shift and a dc linear Stark
shift   of the $\vert F,m_F \ket $ ground state sublevels, violating P but not T.  It is 
quadratic in the parity conserving (PC) atom radiation field interaction, and linear in the static interaction of the
nuclear anapole moment with a dc Stark field. Its physical origin is then  totally different from
the PV light-shift involved in a project for  a measurement of $Q_W$ on a single trapped Ba$^{+}$ ion \cite{for93}. In
the present project, one must apply both a static electric field and a strong laser field, but  
its phase is irrelevant. So it is not necessary to localize the atoms inside a cavity
%, hence,  as to its physical origin, it totally differs from
%the PV light-shift suggested to measure $Q_W$ on a single trapped Ba$^{+}$ ion \cite{for93}. To observe it, one
%must apply both a static electric field and a strong laser field but without any requirement on its phase, 
%hence without any need to localize the atoms inside a cavity, 
unlike the case of previous proposals
\cite{ezh,deM}.

The nuclear spin-dependent PV interaction is responsible not only for a transition electric dipole but also for a
permanent one in an alkali atom ground state, involving the operator  $\vs \wedge \vI$, $\vs$ and $\vI$ being
the spin of the electron and the nucleus respectively.  Thus, in an applied static electric field $\vE$, the nuclear
anapole moment gives rise to the time-independent P-odd, T-even, interaction $V_{ana} = d_I \vE \cdot \vs
\wedge
\vec I \equiv -d_I \,\frac{i}{2} [F^2,\vs\cdot \vE]$, 
where $\vF= \vs + \vI$. This identity results from properties of the Pauli matrices, $\vec \sigma =
2 \vs$. The  commutator appearing in the last expression shows directly that $V_{ana}$ has 
matrix elements only between states belonging to different F multiplets. 
From previously published data
\cite{woo97}, the electric dipole moment $d_I$ for the Cs ground state
can be estimated to $d_I \simeq 2.36(40) \times 10^{-13}$ $\vert e
\vert a_0$ \cite{bou01}.  

We assume that the atom in its ground state is also
perturbed by the  electric field 
of a laser beam of frequency $\omega$
detuned from an allowed transition of frequency
$\omega_0 =\omega -\delta$.
Supposing the beam circularly polarized, we write the associated electric classical field ${\vE}_c(\vr ,t) $:
\bea {\vE}_c(\vr ,t) 
=\frac{1}{\sqrt{2}}\,  {\cal E }\(( \ve (\xi) \, \exp \,( -i \,\omega \,
t +i \, \vk \cdot  \vr) + c.c. \,\)) , \eea
with $  \ve (\xi) =\frac{1}{\sqrt{2}}\, ( {\ve
}_1 +i\,\xi \,{\ve }_2)$, $ \ve_i$  being two orthogonal  unit real vectors normal  to
the wave vector  $ \vk $  of the laser beam and $\xi = \pm1$ defining the helicity. 
 The energy density of the beam  $ \epsilon_0\, {\vE}_c(\vr ,t)^2$
  is then  given by  $\epsilon_0\,{\cal E  }^2 $.
We  write the  quantized form of the radiation electric field,
assuming that  only the laser mode $ \ve (\xi) \, \exp \,( -i \,\omega \,
t +i \, \vk \cdot  \vr )$ is involved in
the  radiative transition. The associated annihilation (creation)  operator
$ a \, ( a ^\dagger) $ is normalized in
such a way that the one  photon state obtained by applying  $ a ^\dagger $
upon the vacuum state $\vert 0 \ket $ has its
energy  $ \hbar  \omega $ localized inside a volume $V$.
If the laser beam is described by a normalized
$N$  photon state $ \vert N \ket $, its energy  density is:
$\epsilon_0\,{\cal E  }^2 = N\, \hbar  \omega /V$. 

The combined atom-field system is described in terms of its eigenstates $\vert i \ket \vert N\ket$,
where the first ket refers to the atomic states and the second to the radiation field states. The  coupling  
of the ground state  to the
nearly resonant excited state via  virtual photon  absorption and emission
gives the dominant contribution to the ac Stark shift, or light-shift, of the $6S_{1/2}$ substates. The
hamiltonian $V_{rad} = -\sqrt {\frac {\hbar \omega}{2 \epsilon_0 V}} (a \; \ve(\xi) \cdot  \vD + h.c. )$ describes
the atom-field dipolar coupling. Its matrix elements between the initial state,
$\vert\, i \ket =\vert n\, S_{1/2}\ket \, \vert N \ket  $, 
and the first excited state $\vert\, f \ket =\vert n^{\prime}\, P_{J}\ket \, \vert N-1 \ket  $ read:
\bea \bra f \vert V_{rad}\vert\, i\ket 
 = -\frac{1}{\sqrt{2}} \,{\cal  E}\, \,\bra n^{\prime}\, P_{J}\vert
\ve (\xi)\cdot \vD \vert n\, S_{1/2}\ket .
\eea
   
 We write the lowest-order non-zero modification to the ground state energy  involving both perturbations 
$V_{ana}$ and $V_{rad}$. It is linear in $V_{ana}$ and quadratic in $V_{rad}$: 
%(where $V_{rad} \ll V_{rad}$) 
\begin{eqnarray}
   E^{(3)}_{\alpha} = \bra \alpha \vert V_{ana} R V_{rad} R V_{rad} \vert \alpha \ket + \hspace{20mm}
\nonumber \\
  \bra \alpha \vert V_{rad} R V_{ana} R V_{rad} \vert \alpha \ket +\bra \alpha\vert V_{rad} R V_{rad} R
V_{ana}  \vert \alpha \ket.
\end{eqnarray}
In this general expression,  the hyperfine and the Zeeman structures are now included in the description of the
atom-field states: 
$\vert \alpha \ket= \vert nS_{1/2}; F, m_F\ket\vert N\ket$, 
  the resolvent operator $R=(E_n -H)^{-1}$ involves the
unperturbed hamiltonian $H$ and the ground state energies $E_n$ of the combined atom-field system. 
As it will become clear later on, the
condition $\vert \Delta W / \hbar \delta \vert  \ll 1$ has to be satisfied in a realistic experiment. Therefore, 
the second term in Eq.(3) can be neglected because its energy denominator $\vert \hbar \delta \vert^2$ is much
larger than  those of the two other terms, $\Delta W \vert \hbar \delta \vert$.  
\begin{equation}
 E^{(3)}_{\alpha} = \frac{(F-F^{\prime})}{\Delta W}\bra \alpha \vert V_{ana}\vert
\alpha^{\prime}\ket
\bra \alpha^{\prime}
\vert  V_{rad} R V_{rad} \vert \alpha \ket + h.c. 
\end{equation} 
Here the state $\vert \alpha ^{\prime} \ket $ differs from $\vert \alpha  \ket $ just by the change of $F,m_F$ 
into $F^{\prime}\not = F, m_F^{\prime}$. The last factor is the second-order modification to the
ground  state energy arising from its coupling with the radiation field. First, we limit ourselves to the effect of the
laser detuning 
with respect to the single $P_{1/2}$ excited state.    
In this case, $V_{rad}$  operates between $S_{1/2}$ and $P_{1/2}$, {\it i.e.} two
electronic states of angular momentum $J=\frac{ 1}{2}$, and the atomic part of the operator involved  can 
be simplified to    
$V_{rad} = \hbar \Omega_1 \vec \sigma \cdot \hat \epsilon$, where $ \Omega_1=  d \; {\cal
E}/\hbar\sqrt{2} $ is the Rabi frequency, and $d$ the  $\Delta m_s = 0 $ electric dipole matrix element. The
operator R can be replaced by the projection operator
$ {\cal P}_{P_{1/2}}= \sum_m \vert P_{1/2}, m\ket \bra P_{1/2}, m\vert$ divided by $\hbar \delta$.  A change
of $F$ into $F^{\prime}
\not = F$ can be produced only by  the vector part of the tensor operator
$V_{rad}\, {\cal P}_{P_{1/2}}\, V_{rad} $,  which is just $\hbar \Omega_1^2  \; \vec \sigma \cdot  
\xi  \vk $, 
if we introduce the angular momentum of
the photons $\xi  \vk= i (\hat \epsilon \wedge   \hat \epsilon^{*}) \hbar$.  
Then, we can substitute the expressions of $V_{ana}$ and $V_{rad}$ into Eq.(4), to obtain the anapole
energy shift of a $\vert F,m_F \ket$ substate:
\begin{equation}
h \, \Delta  \nu_{F,m{_F}}^{ana} =  
\frac{ d_I  \;  \;  \Omega_1^2} { \Delta W \;\, \delta} \times \nonumber 
\end{equation}
 \vspace{-6mm}$$  \sum_{F^{\prime},
m^{\prime}_F}  (F^{\prime}-F)\bra F,m_F\vert \vE \cdot \vec \sigma \wedge \vI  \vert F^{\prime},
m^{\prime}_F\ket 
\bra F^{\prime}, m^{\prime}_F\vert \vec \sigma \; \cdot \xi  \vk \vert F, m_F\ket \; .  $$
 Formally,  the summation over $F^{\prime}$can be extended to all states, although only the states
$F^{\prime}\not =F$ do contribute. By using a closure relation, we obtain the compact expression 
$\bra F, m_F\vert \vE\cdot \xi \vk \wedge \vI \vert F, m_F\ket $.
Taking as quantization axis 
the direction defined by  the applied magnetic field $\vB$, 
we arrive at the final expression for the anapole shift: 
 %If, now, we inroduce the direction of the
\begin{equation}
h \, \Delta \nu_{F,m{_F}}^{ana} =  
 2(F-I)(\xi \hat k \wedge \hat E \cdot
\hat B )\frac{d_I E \; \; \hbar\Omega_1^2} { \Delta W \; \delta}   \bra F, m_F \vert I_z\vert F, m_F \ket, 
\end{equation}
where $\hat B$, $\hat k$ and$\hat E $ are respectively unit vectors parallel 
to $\vB $, $\vk$ and $\vE$.  
The $I_z$ matrix elements take the form $\lambda_F m_F$, in Cs $\lambda_F= \frac {7}{8}$ and
$\frac{9}{8}
$ for F=4 and 3 respectively. 
We note that shifts of opposite signs are predicted for the two $\vert F=I\pm
\frac{1}{2},m_F \ket$ states, hence both contribute constructively to the frequency shift of the $\vert \Delta F
\vert =1,
\Delta m_F = 0$ transitions which, thus, scales as $2 m_F$.  
If $\vB$ defines the quantization axis (see below), then, its magnitude has no
effect on the anapole shift at the second order in $V_{rad}$  where Eq.6 is only valid.
 
The presence of the pseudoscalar quantity $\xi \hat k \wedge \hat E \cdot \hat B = \chi$ is the mark of parity
violation \cite{noteA}. If the two fields and $\xi \vk$ form a rectangular trihedron,   
 its chirality $\chi$ takes the value $\pm 1$. 
%if  the two $\vE$ and $\vB$ fields and  the direction of the circularly polarized beam form a
%rectangular trihedron. 
Reversals of  $\xi$, $\vE$, $\vB$ and
$m_F$ would be the indispensable operations used to identify the anapole frequency shift. They are all the
more helpful since,  without special care, this shift may appear superposed on a larger    
PC light-shift associated with the scalar part of the operator $V_{rad}\, {\cal P}_{P_{1/2}}\, V_{rad}$. 
\begin{figure}
\vspace{-32mm}
\centerline{\hspace{10mm}\epsfxsize=100 mm \epsfbox{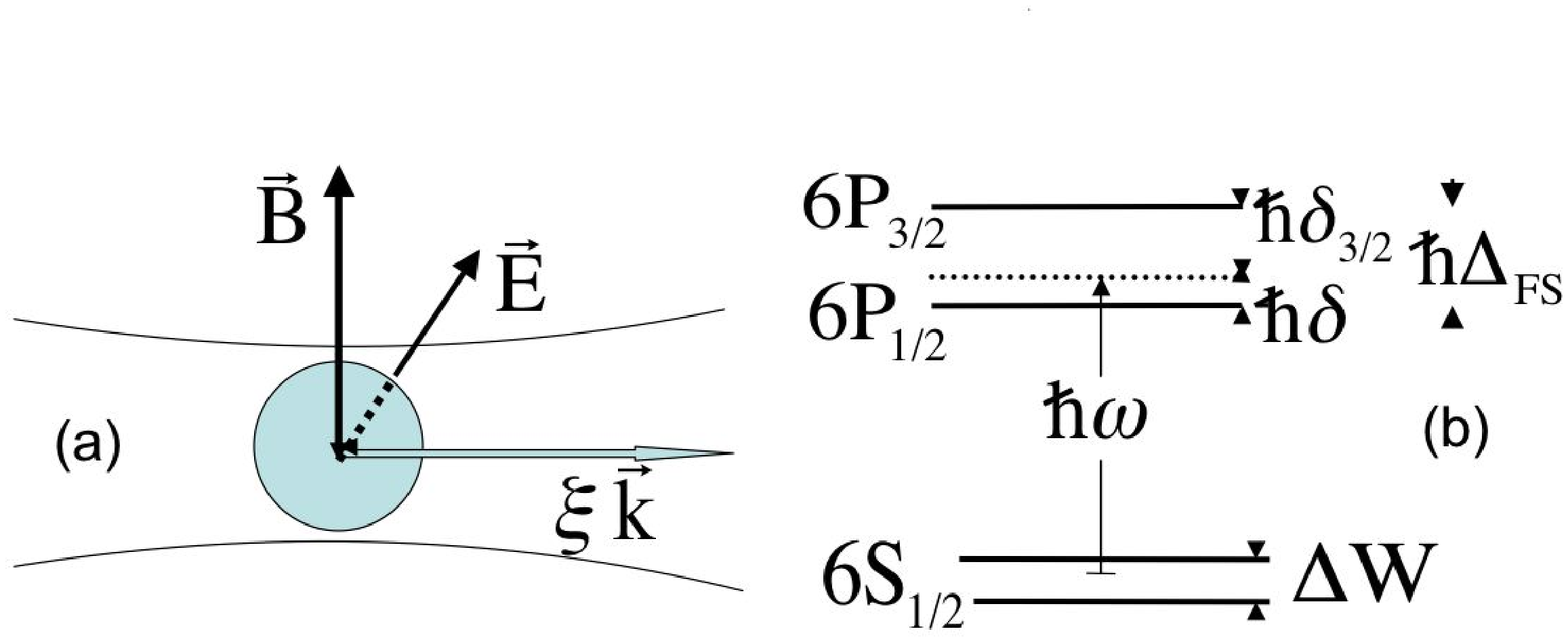}}
\vspace{-15mm}
\caption{(a) Scheme of the interaction region located close to the apex of the cold atom fountain. (b) Schematics
of the relevant Cs atomic levels  and laser detunings for $\delta_{3/2}/\delta <0$.  }
\vspace{-5mm}
\end{figure}
Due to the slight detuning difference between the two F states, the scalar light-shift of a $\vert F,
m_F\ket$ state depends  on $F$ to first order in $\Delta W/ \hbar \delta$. Thus, the well known shift
of the hyperfine  splitting is 
$- \frac {  \Delta W \Omega_1^2 }{h \delta^2}$, whatever the sign of $\delta$. As for the contribution of the
$P_{3/2}$ state,  when its  frequency detuning $\delta_{3/2} $ is comparable to $ \delta$,   it always reinforces
that of $P_{1/2}$.

Let us, now, consider the relative contributions to the anapole
and the PC light-shifts coming from the vector parts of the operator $V_{rad}R\,V_{rad}$ associated with both
$P_J$ states. The basic property of the projection operators implying  
${\cal P}_{P_{3/2}} = {\bf 1} - {\cal P}_{P_{1/2}}$, where ${\bf 1}$ is the unit operator, allows us to conclude 
that they are just opposite. 
Once weighted by the energy denominators, $\hbar \delta_{1/2} \equiv \hbar \delta $ and $\hbar
\delta_{3/2}=- \rho \; \hbar \delta $, we see that, provided that the detunings be of opposite signs  
 ($\rho>0$, Fig.1b),  both contributions add together and reinforce  both the
anapole shift (Eqs.(4) to (6)) and the vector PC light-shift  $\xi \frac{\hbar \Omega_1^2}{\delta}$ by the factor
$(1+\rho)/\rho  $. (Note that the parameter $\rho$ is simply related to
$\delta$ by $(1+\rho) \delta  = \Delta_{FS}$, the fine structure interval and appears just for writing
convenience). As shown in \cite{jdr72}, the light shift induced within the $6S_{1/2}$ sublevels by a circularly
polarized beam can be described with the help of a fictitious magnetic field 
$\vB_{ls}= \xi \vk \frac{\hbar \Omega_1^2 }{\vert \delta \vert \; \mu_B} \frac{1+\rho}{\rho}$. 
This provides a useful bench-mark to compare Zeeman and light shifts. As an example, if  $\vk \cdot \vB=0$,
the tilt of the quantization axis induced by $\vB_{ls}$ leads a correction to the anapole shift of Eq.6 which
reads to lowest order: $\frac{\delta \Delta \nu^{ana}}{\Delta \nu^{ana}}\approx -\frac{1}{2} (\frac
{B_{ls}}{B})^2$ for $B \gtrsim B_{ls}$.  
For $ \vk \cdot \vB=0$ the
$\xi$-dependence of the hyperfine splitting also cancels.

Before discussing the anapole shift magnitude, let us indicate how the measurement might be
conducted in a fountain clock type experiment. A ball of cold atoms in an $\vert F,m_F \ket$ state is launched
vertically along the applied magnetic field. It passes through a microwave cavity and gets prepared  in a coherent
superposition of the
$F,m_F$ and $F',m'_F=m_F$ states described by the wave function
$\vert \psi \ket = (\vert F,m_F\ket \vert N \ket + \vert F',m_F\ket \vert N \ket)/\sqrt{2} $. After rising a 
distance $L$ above the cavity centre, the atoms reach the interaction
region, located close to the apex of their trajectory.  The dc field
$\vE$, the polarized laser beam $\xi \vk$, and $\vB$ form a rectangular trihedron (Fig.1a). Due to reversal of
the atomic velocity, the very small motional magnetic field cancels.  The beam waist should be chosen larger  or
equal to the atomic ball radius,
$\sim 5
$ mm. The atoms falling back reach the  detection region. As
 usual in a fountain clock the measurement relies on the  phase shift of the atomic superposition
accumulated during the total time $\tau_f $ elapsed since its preparation, $\int \phi(t) dt$. Since the interaction
lasts for a limited duration $\tau_i$, we expect the variation of the phase shift for the interaction on and
off $(\Delta \nu^{ana}_{F,m{_F}} - \Delta \nu^{ana}_{F',m{_F}})$, to be reduced by the factor $\tau_i/ \tau_f
\simeq \sqrt {2w/L}$, typically 0.18. 

Up to now, $\Omega_1$ has appeared as a free parameter. However, it must be kept in mind that it
cannot be chosen arbitrarily large : from the coupling induced between the ground and
excited states by the radiation field results an instability of the atomic sample crossing the
interaction region, hence a signal loss. The decay rate acquired by the 6S atoms is easily computed by
perturbation theory in
the limit $\Gamma_P{_J} \ll \Omega_1$, 
$\Gamma_S=  
\frac{\Omega_1^2}{\delta^2} \Gamma_{P_{1/2}} \kappa$, 
with $\kappa= 1+ \frac{R^2_{Rabi}}{\rho^2} \;\Gamma_{P_{3/2}}/\Gamma_{P_{1/2}}$, where $R_{Rabi}$ 
is the ratio of the Rabi frequencies for $P_{3/2}$ and
$P_{1/2}$. Using numerical values from 
\cite{tan98}, $R^2_{Rabi} = 1.98$ and $\kappa = 1+2.27 /\rho^2$   for  Cs.    
The condition to avoid excessive signal loss is $\Gamma_S \tau_i \leq 1$. 
Assuming this limit just reached, the anapole shift of the 
$F,m_F \rightarrow F+1, m_F$ transition angular frequency becomes: 
\begin{eqnarray}
&& \hspace{-8mm}2\pi \, \Delta \nu_{m_F,m_F}^{ana}=2 \, \chi \, m_F \frac{1+\rho}{\rho} \, \frac {d_I
E}{\Delta W } \; \frac{\Omega_1^2}{\delta}\;
\frac {\tau_i}{\tau_f},  \\
=&&\hspace{-3mm}2 \, \chi \, m_F  \frac{ (1+\rho)}{\rho  }\,\frac {d_I E}{\Delta W \tau_f }
\; \frac{\delta}{\kappa\Gamma_{P_{1/2}}}, \hspace{1mm} {\rm for} \hspace{2mm}
\Omega_1^2= \frac {\delta^2}{\kappa \Gamma_{P_{1/2}} \tau_i} \,.\hspace{3mm}
\end{eqnarray}
 Eq.(8) shows that, due to the constraint imposed on the laser
intensity, it is important to choose $\hbar \delta /\Delta W \gg 1$, as announced earlier. 
 Using $\rho$ as the sole free parameter,  it is easily found that $\Delta
\nu_{m_F,m_F}^{ana}$ as a function of $\rho$ presents a maximum for $\rho = \sqrt{2.27} $, leading to
$\kappa = 2$.   

Inserting in Eq.(8), $E = 100$ kV/cm, $d_I E /h = 30 $~mHz, $\delta/2\pi= 6.65  $~THz, $\Delta
W/h = 9.2
$~GHz, $\tau_f = 0.5 $ s, $\tau_i/\tau_f = 0.18$, $\Gamma_{P_{1/2}} ^{-1}= 35  $~ns and $\rho = 1.50$,
we obtain for Cs:
\begin{eqnarray}
\Delta \nu_{3,3}^{ana}&=&7.6 \; \mu{\rm Hz,}  \\
{\rm and}\hspace{5mm}   \Omega_1/2\pi &=&2.9 \; {\rm GHz,} \hspace{5mm} B_{ls}= {\rm 0.24
\; G.}    
\end{eqnarray} 
 For a beam radius of 5 mm, the laser power should be $\approx $1 kW. This can be achieved inside a
Fabry-Perot cavity with a reasonable input power tuned at 877 nm \cite{noteB}. 
 The condition of adiabatic passage is largely satisfied by the atom-radiation field interaction.
One might object that the magnitude chosen
for $E$ is large. However, still larger fields (450 kV/cm) have been achieved before, for quadratic
Stark effect measurements with heated glass electrodes \cite{gou76}.  The quadratic differential
dc Stark shift \cite{cla98}, $\Delta \nu^{dc} \simeq 10$ kHz,  and the smaller scalar light-shift,
$\Delta \nu^{ls} = -(1 + \frac{R_{Rabi}^2}{\rho^2})  \frac{\Delta W \Omega_1^2}{h \; \delta^2}
\frac{\tau_i}{\tau_ f} $, could be eliminated via the fastest parameter reversals among the set $m_F, \xi, E, \,B$. 
\begin{figure}
\vspace{-18mm}
\centerline{\hspace{-8mm}\epsfxsize=100 mm \epsfbox{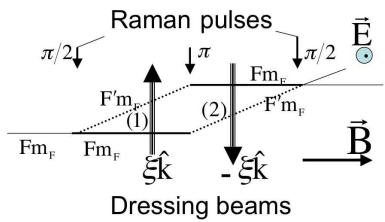}}
\vspace{-25mm}
\caption{%Both paths of the interferometer cross two interaction regions having chiralities of opposite signs.
Scheme of the interferometer. Both paths cross two dressing beams having opposite photon angular momenta, 
$\xi \vk$.  }
\vspace{-5mm}
\end{figure}
 
It is clearly of interest to perform a calibration of the anapole shift allowing for a determination of the basic
parameter $d_I E$ free of the uncertainties coming from the laser power and the exact
geometry. We suggest using  the PC scalar light-shift obtained by
offsetting  sequentially  the laser detuning by a fractional amount $\pm \eta$ from one side of its best 
value  to the other. In this case, $\Delta \nu_{\eta-odd}^{ls} =  \Omega_1^2 
\frac{\tau_i}{\tau_f}\times \, 0.84 \,  \frac{\Delta W}{h \, \delta^2} \; \eta$ is the light-shift odd in the $\eta$
reversal for cesium, while $\Delta \nu^{ana}$, measured at its optimum, is not affected to first order in $\eta$. In
the ratio, 
\begin{equation}
{\cal{R}}(\chi)=\frac{\Delta \nu^{ana}_{3,3}}{\Delta\nu_{\eta-odd}^{ls}} = 12. \chi \,  \frac{d_I E }{\Delta W}
\,\frac{\hbar
\delta}{\Delta W} \,\eta^{-1}, 
\end{equation}
which bears the PV signature $\chi$,  
$\Omega_1^2 \tau_i/\tau_f$ is eliminated and  the only unknown quantity is just that we need,
$d_I E/\Delta W$. 
The intensity-independent left-right asymmetry ${\cal{R}}(+1)-{\cal{R}}(-1)$ affecting the hyperfine transition
frequency is definitely the quantity to be measured. It amounts to $ \approx 1.2 \times 10^{-3}$ for $\eta  =
0.45
\times 10^{-4} $ (laser frequency excursions of 3 MHz).   
The precision will depend on the magnitude of $\eta$, and the
frequencies of the parameter reversals adjusted for limiting the effect of harmful drifts. The most severe of these 
are likely to be  the $\vB$ drifts.  Since the 3,0
$\rightarrow $ 4,0 transition is insensitive to the anapole shift, measurements should be
performed on a  first-order Zeeman shifted transition, as in the case of  the LI tests \cite{cla}. 
 A dual Cs-Rb fountain, already used for a fundamental physics test \cite{Syr02}, could be very helpful, since
frequency comparisons should reduce considerably the magnetic noise.  
 
Another approach would be to 
use a cold atom interferometer. In the case where the wave
packets are separated and recombined by using a $\pi/2-\pi - \pi/2$ sequence of Raman pulses \cite{Chu},
both paths can be made to cross two interaction regions having chiralities of opposite signs (Fig.2). In the phase
difference
$\Phi$ between the two paths, it is easily verified that the Zeeman shift cancels while the anapole shift
contribution doubles. The same property holds for the calibration signal if $\eta$ is applied with opposite
signs on the two dressing beams. An estimate gives
$\Phi \approx 10 ~\mu$rad for  $\tau_f=100$~ms, $\tau_i/\tau_f=0.2$,
$\Omega_1^2$ cut off by 2 for keeping $\Gamma_S$ as all other parameters the same. 
Since no absolute measurements is required, this very specific phase shift looks accessible  in the present
state of the art.

In conclusion, we have shown the existence of a linear atomic Stark shift. with a chiral  character providing a
signal  for the  nuclear anapole moment. We have exhibited a concrete example  where several powerful
techniques  developed   recently in atomic physics for high precision frequency determinations appear  as highly
valuable tools for precise atomic PV measurements. The effect involved presents a certain similarity with the  
PV NMR frequency shifts expected  in enantiomer  molecules
\cite{mol}. The chiral configuration defined by  $\xi \vk, \vE, \vB$,
is  playing here the role of the  chiral 
arrangement of the atoms inside the molecule,  
but with the advantage of  providing  numerous reversals.  Owing to the absence of any
contribution due to the weak nuclear charge, one could thus obtain a rather
direct measurement of a PV static property of the nucleus. The method
could be extended to other atoms, in particular to strings of (possibly radioactive) isotopes and would
provide valuable information about nuclear parity violation.

The author is grateful to C. Bouchiat, J. Gu\'ena and M. D. Plimmer for helpful remarks. Laboratoire
Kastler Brossel is a Unit\'e de Recherche de l'Ecole Normale Sup\'erieure et de l'Universit\'e Pierre et Marie Curie,
associ\'ee au CNRS (UMR 8552).


\begin{thebibliography}{60}
\bibitem{noteA}
 For a linearly polarized laser beam, the pseudoscalar involved is $(\hat \epsilon\cdot\hat B)( \hat
\epsilon \cdot \hat B \wedge \hat E $), but the  effect cancels unless the 
$P_J$ state hyperfine interaction is taken into account. This reduces its magnitude by a factor
$E_{hf}/\delta \approx 10^{-4}$ .
\bibitem{noteB}
Choosing $\rho=R^2_{Rabi}$ cancels the radiative potential, and its possible perturbative effect on the atomic
motion, at the price of a reduction of $\Delta \nu^{ana} $ of only 4$\%$.
\end{thebibliography}
 \end{document}